\documentclass{article}
\usepackage{arxiv}
\usepackage[utf8]{inputenc} 
\usepackage[T1]{fontenc}    
\usepackage{hyperref}       
\usepackage{url}            
\usepackage{booktabs}       
\usepackage{amsfonts}       
\usepackage{nicefrac}       
\usepackage{microtype}      
\usepackage{lipsum}
\usepackage{graphicx}
\usepackage{notoccite}
\usepackage{subcaption}
\usepackage{booktabs}
\usepackage{graphicx}
\usepackage{adjustbox}
\usepackage{wrapfig}

\usepackage[hang,flushmargin]{footmisc}

\usepackage[style=numeric,sorting=none]{biblatex} 

\graphicspath{ {./images/} }

\title{Evaluating Twitter's Algorithmic Amplification of Low-Credibility Content: An Observational Study}

\author{
  Giulio Corsi \\
  Leverhulme Centre for the Future of Intelligence\\
  University of Cambridge\\
  \texttt{gc540@cam.ac.uk} 
}

\begin{document}
\maketitle
\vspace{10pt}
\begin{abstract}

Artificial intelligence (AI)-powered recommender systems play a crucial role in determining the content that users are exposed to on social media platforms. However, the behavioural patterns of these systems are often opaque, complicating the evaluation of their impact on the dissemination and consumption of disinformation and misinformation. To begin addressing this evidence gap, this study presents a measurement approach that uses observed digital traces to infer the  status of algorithmic amplification of low-credibility content on Twitter over a 14-day period in January 2023. Using an original dataset of $\approx$ 2.7 million posts on COVID-19 and climate change published on the platform, this study identifies tweets sharing information from low-credibility domains, and uses a bootstrapping model with two stratifications, a tweet's engagement level and a user's followers level, to compare any differences in impressions generated between low-credibility and high-credibility samples. Additional stratification variables of toxicity, political bias, and verified status are also examined. This analysis provides valuable observational evidence on whether the Twitter algorithm favours the visibility of low-credibility content, with results indicating that tweets containing low-credibility URL domains perform significantly better than tweets that do not across both datasets. Furthermore, high toxicity tweets and those with right-leaning bias see heightened amplification, as do low-credibility tweets from verified accounts. This suggests that Twitter’s recommender system may have facilitated the diffusion of false content, even when originating from notoriously low-credibility sources.
\end{abstract}

\section{Introduction}
The emergence of social media platforms has brought about a significant transformation in global patterns of information dissemination and consumption, as a large number of internet users now rely on these channels as their primary sources of information acquisition \cite{pentina2014,shearer2017,walker2021,cinelli2021}. The rapid growth in social media membership, and consequently of digital traces circulating in these platforms, have been accompanied by a progressive rise in the importance of artificial intelligence (AI) based recommender systems, content pre-selection, ranking and suggestion systems used to customise users’ online experiences \cite{anandhan2018,thorburn_2022}.

The integration of AI-based recommender systems into social media platforms has led to a fundamental shift in the way users consume and interact with online information \cite{santos2021}, significantly increasing the level of automated content curation while limiting users’ freedom of independent content discovery \cite{pariser2011}. This paradigm shift towards the machine-learning based hyper-personalisation of social media content raises concerns regarding potential impacts on the quality and diversity of information available to users, with clear implications for the integrity of knowledge acquisition processes. Several recent studies have analysed these risks, concluding that engagement-based recommender systems - which form the majority of recommendation engines currently deployed within social media platforms - may be prone to bias \cite{islam2019,bhadani}, user-manipulating behaviour \cite{burns2023}, the creation of echo chambers \cite{anandhan2018,alatawi2021}, and to the amplification of false or misleading content \cite{kaiser2021,giansiracusa2021}.

Despite their status as critical infrastructure of social media platforms - and arguably of information circulation at a societal level - the internal architectures and practical functioning of recommender systems remain only superficially understood \cite{leerssen2020}, and while several platforms have previously released white papers with information on their functioning \cite{liu2022,zhao,lada2021}, limited evidence exists on the characteristics that guide their deployment. This is also the case for Twitter, (Now X Corp.) which recently made parts of its recommender system public, providing a window into the functioning of a social media content suggestion system \cite{twitteralgo}. However, while this release does provide new information on the system’s architecture, perhaps the most central part of the system - a ‘heavy ranker’- deep neural network used to make recommendation predictions \cite{wang2021} cannot be replicated with currently available information, limiting the possibility of testing the behaviour of this recommender system. This lack of evidence is a clear obstacle towards evaluating the magnitude of any form of algorithmic bias in content suggestion, and in particular, towards understanding whether, in their drive to maximise user engagement with a platform, recommender systems are acting as significant drivers of the diffusion of online disinformation and misinformation. This is crucial, as understanding how false and low-credibility content propagates within social media platforms is key to improving the safety of societal information commons.  

In an effort to address the existing evidence gap, this study introduces a measurement approach that leverages existing digital traces to empirically observe - analysing recommendation outcomes through impression counts -  the state of the promotion of low-credibility content on Twitter in a two week period in January 2023. The motivation for choosing Twitter as the object of this study is threefold: First, Twitter has a large user base with global reach, with a monthly user base of approximately 450 million users and over 300 million daily tweets \cite{Pfeffer}, making it one the largest and most influential social media platforms globally. Second, Twitter has been often criticised for platforming and amplifying extremist content, disinformation and misinformation \cite{bovet2019, kouzy2020}, and studying its recommender systems may provide additional insights into how false information spreads on the platform. Third, Twitter was, at the time of data collection, the only platform which provided data on impressions through its API. While Twitter’s recommender system has undergone significant changes since January 2023, this study aims to offer a snapshot of the recommender system’s behaviour at a crucial juncture in the platform's evolution, before the recent system-wide overhaul. Although the findings may not directly translate to the current recommender system, they may provide unique insights into a pivotal transitional moment, revealing baseline tendencies in how the system influenced the propagation of low-credibility content.

To map the behaviour of Twitter’s recommender system, this study uses an original dataset of $\approx$ 2.7 million tweets discussing COVID-19 and climate change published on the platform in a 14-day period in January 2023, and extracts tweets containing information from low-credibility domains and high-credibility domains \cite{lin2023}, testing whether low-credibility information gets amplified visibility on the platform. Data on impressions - defined in the API documentation as the count of how many times a Tweet has been seen \cite{twitterengineering} - was initially made available through the Twitter API in January 2023 and, as a passive metric of how many users have been exposed to a tweet, provides a powerful window into the content that the Twitter recommender system tailors to users. Through this process, it is possible to estimate whether tweets sharing information from low-credibility domains generate exceptional impressions, which would point towards a recommendation bias towards low-credibility content, as well as a general lack of functioning integrity signals \cite{thorburn_2022}. By analysing the visibility of low-credibility content - which can be used as a baseline for false or misleading information - this research seeks to provide insights into the dynamics that drive user exposure to false or misleading information, with potential implications for social media platforms and the broader digital information ecosystem.

\section{Data and Methods}
\subsection{Data}
\subsubsection{Twitter Data}

\begin{wrapfigure}{r}{0.5\textwidth}
    \centering
    \includegraphics[width=0.48\textwidth]{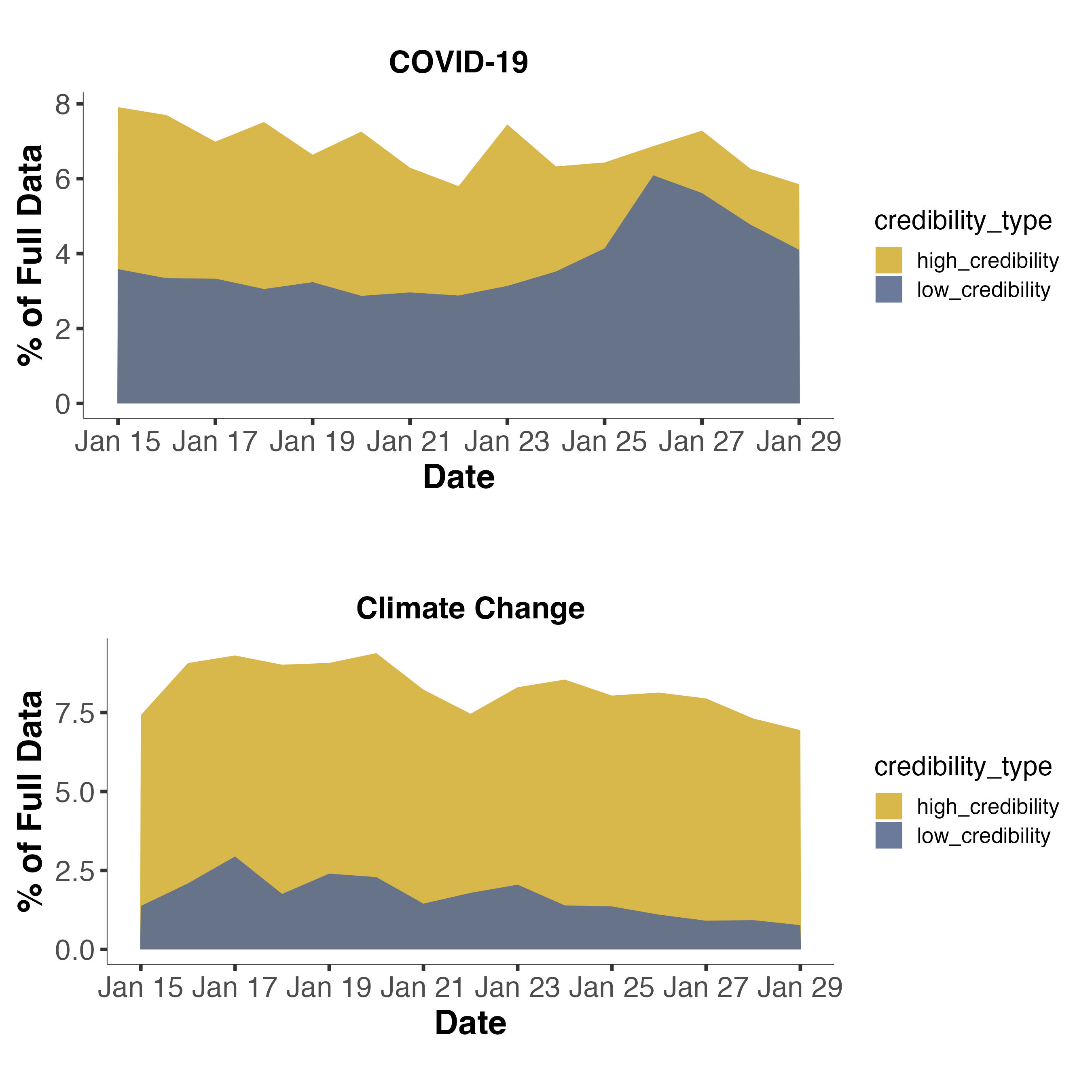}
    \caption{Distribution of tweets with low-credibility and high-credibility domains as a percentage of the full data in each dataset under analysis}
    \label{fig:frequency}
\end{wrapfigure}

The data utilised in this study was collected from the Twitter API V2 in a 14-day time period between January 15th to January 29th, 2023. Given the research objective of assessing whether the Twitter recommender system amplifies the visibility false or misleading information, this work uses data from discussions on COVID-19 and climate change, two debates that are often considered publicly divisive and subject to a significant circulation of false content \cite{treen2020,brennen2020,graham2020}. Data discussing these topics was collected through a keyword search using the R package AcademictwitteR \cite{Barrie}, which at the time of collection, was chosen for its exclusive capability of collecting data on a tweet’s impressions. The data was collected at regular intervals ensuring similar uptime for each day of publication, and the resulting dataset comprises a total of $\approx$ 2.1m original tweets on COVID-19, and $\approx$ 600k original tweets on climate change. Data collected from the Twitter API is quite granular, and is made up of 21 variables, encompassing several tweet-level and user-level variables, such as engagement metrics, verification statuses, locations and users’ profile images. 

\subsubsection{URL Domains Classification}
In order to extract tweets that are likely to contain false or misleading information, this study relies on the use of URL domain credibility ratings, which is often used in the literature on the study of disinformation and misinformation \cite{pierri2022,falkenberg2022,resnick2018}. While several datasets domain credibility exist, such as NewsGuard and IffyNews, this study identifies information reliability scores using the aggregate reliability scores from \textcite{lin2023}, which used principal component analysis to produce aggregate scores derived from the major available rating sets. This dataset provides credibility scores for 11,520 domains, where 0 represents the lowest credibility, and 1 represents the highest credibility.  In line with the values used by major credibility ratings providers such as Newsguard, we then consider tweets with a credibility score lower or equal to 0.4 to be low-credibility and tweets with a credibility score equal or higher than 0.6 to classify as high-credibility \cite{newsguard}. This process results in a total of 87,769 tweets from low-credibility sources, and 187,643 tweets from high-credibility sources, with low-credibility domains present in 3.77\% of tweets on COVID-19 and 1.69\% of tweets on climate change. The most common low-credibility and high-credibility domains for both datasets are shown in Table 1. 

\begin{table}[h]
\centering
\renewcommand{\arraystretch}{1.5}
\begin{tabular}{lll|lll}
\hline
\multicolumn{3}{c|}{\textbf{COVID-19}} & \multicolumn{3}{c}{\textbf{Climate Change}} \\
\textbf{Key} & \textbf{Domain} & \textbf{Frequency} & \textbf{Key} & \textbf{Domain} & \textbf{Frequency} \\ 
\hline
\textbf{High Credibility} & theguardian.com & 7414 & \textbf{High Credibility} & theguardian.com & 6409 \\
 & reuters.com & 5364 &  & nytimes.com & 1275 \\
 & washingtonpost.com & 3056 &  & washingtonpost.com & 1239 \\
 & cnn.com & 2868 &  & independent.co.uk & 925 \\
 & msn.com & 2499 &  & weforum.org & 812 \\
\textbf{Low Credibility} & rumble.com & 13911 & \textbf{Low Credibility} & breitbart.com & 925 \\
 & thegatewaypundit.com & 9614 &  & thegatewaypundit.com & 862 \\
 & theepochtimes.com & 5645 &  & rumble.com & 860 \\
 & expose-news.com & 5230 &  & dailymail.co.uk & 850 \\
 & zerohedge.com & 5047 &  & zerohedge.com & 705 \\
\hline
\end{tabular}
\vspace{10pt}
\caption{Distribution of the five most prevalent high-credibility and low-credibility domains for both datasets}
\end{table}

\vspace{-5pt}

\subsection{Measuring Amplification}
\subsubsection{Baseline Amplification Benchmark}

Measuring recommender-driven amplification is a notoriously difficult task, which requires clearly defined objectives and robust benchmarks to identify potential patterns of amplification. In this study, amplification is defined as a condition where tweets with similar characteristics drawn from two different groups - low-credibility and high-credibility - exhibit a significant difference in the outcome variable, which is the number of impressions obtained. Amongst the available metrics currently provided by the Twitter API, data on impressions is the most suited for the study of recommender-based amplification, as it is the only available passive metric of exposure, measuring how many people a certain tweet has been shown two irrespective of voluntary engagement. As users have no control on impressions, this metric is expected to be effective in characterising the behaviour of the recommender system \cite{thorburnetal2023} .
To produce a clear measure of amplification, it is therefore important to establish a robust benchmarking procedure to compare the two samples under analysis. For this purpose, this study compares the two previously described samples of high-credibility and low-credibility tweets through bias-corrected and accelerated bootstrapping (BCa), where the mean difference between the two samples is measured across 1000 randomly resampled iterations while incorporating a bias-correction factor in the resampling process \cite{efrontibi94, jung2019}. The number of iterations used ensures that the mean difference between the two samples is based on a substantial number of comparisons, reflecting a real difference in impressions. Further, as a non-parametric statistical approach, bootstrapping requires fewer assumptions about the distribution of the data to hold, which makes this approach particularly suited for the study of social media data. 

However, a simple bootstrapping benchmark is not sufficient to reliably determine whether a sample consistently received more impressions than the other, as it neglects potential user-level and tweet-level factors that could influence the number of impressions. To remedy this shortcoming, the baseline benchmark bootstrap comparison is performed with two stratifications, resampling the data by a tweet’s engagement level and the user’s number of followers. Engagement was selected as a baseline stratification variable as engagement-based recommender systems are known to highly value a tweet’s engagement performance \cite{narayanan2023}, and high-engagement tweets are likely to be shown more than low-engagement tweets. Followers count was selected as a baseline stratification variable as, as a networked recommender system, the number of followers of a tweet creator is likely to have a non-negligible impact on how many people are exposed to a tweet. While these two variables alone may not account for the entirety of tweet-level and user-level factors that are likely to influence impressions, adding an excessive number of stratification variables with limited explanatory power is likely to be counterproductive, as it would significantly reduce the number of matched samples. Considering these limitations, stratifying the baseline benchmark by levels of engagement and followers appears the most effective strategy to maximise the accuracy and validity of the results. 

As both engagement and followers count are discrete variables with a large range of values, the complexity of these variables is reduced by assigning the data to discrete clusters using quantile based discretization, an approach that allows for a grouping of the data into similar-sized buckets based on quantile rankings \cite{dougherty}. This approach was tested alongside more traditional approaches to clustering such as HBDSCAN and K-means clustering, and consistently provided a more effective grouping of the data. Following an exploratory analysis of the distribution of both variables, the arbitrary numbers of discrete groups to be identified is set to 4, a number that preserves the original variability of the data without placing undue restrictions on the bootstrapping process, producing a total of 16 combinations of strata of engagement and followers clusters. To guarantee consistent results in the bootstrapping stage, quantile-based discretization is applied to the combined datasets of low and high credibility data for each distinct dataset under investigation.

\subsubsection{Additional Stratification Variables}
After developing a method to compute the baseline level of amplification across the two datasets, we can add further individual stratification variables to test the influence of additional grouping variables across subgroups. For this purpose, each additional stratum is separately added to the baseline benchmark, computing any difference between to the baseline amplification of amplification after the addition of a new stratification variable. At this stage, we test amplification across three additional stratification variables: toxicity scores, political bias and verified status. 

Toxicity scores are obtained through the Perspective API by Jigsaw \cite{lees}, which leverages a machine learning model trained on millions of Wikipedia comments to predict how likely it is that an input text will be perceived as toxic by a reader. Like tweets, Wikipedia comments are largely short and informal, making this model quite suited for the analysis of Twitter data \cite{saveski,cuthbertson}. The Perspective API model produces a toxicity score ranging from 0 to 1 for each input tweet, with 0 having a null probability of being found toxic, and 1 having a high probability of a text being perceived as toxic. To avoid creating an excessive number of categories during the stratification process, the toxicity scores obtained from the Perspective API are used to create 3 clusters of toxicity levels with k-means clustering, allowing for the stratification of our data according to the degree of language toxicity. 

Further, the political bias of the URL domains under analysis is obtained, similarly to the credibility scores, by annotating data with the GPT-4 API. While the use of large language models in data annotation tasks is a new phenomenon, recent literature has extensively analysed the performance of GPT 3.5 and GPT-4 in data labelling tasks, including political stance identification, with both models exhibiting high accuracy \cite{petter,gilardi2023}. To maximise the usability and interpretability of the data, for this task, the model is asked to classify the political bias of the input domain into one of five categories: far-left, left, no bias, right and far-right. The model is also prompted to return a value of -1 whenever it does not have information on a domain, or if the domain is non-political. Through this process, we can identify 5,596 political domains - around 10\% of all domains in the data - which are then used as a stratification variable during bootstrapping to assess whether political bias has an influence on the amplification of low-credibility content. Furthermore, it should be noted that far-left sources were not identified in high-credibility domains, and to maximise the comparability, the data was grouped in two categories, right-bias and left-bias. 

The third and last stratification variable used in this work is a user’s verified status, which is used to assess whether the latter is used to amplify low-credibility content on Twitter. Data on users' verified status, a binary variable with values of  'true' for verified accounts and 'false' for unverified ones, was obtained via the Twitter API. However, it's crucial to clarify that this data refers to the legacy verified status - a verification badge Twitter formerly awarded to prominent users as a safeguard against impersonation. As of November 2022, Twitter began phasing out this legacy verification in favour of Twitter Blue, a paid subscription service that enables users to purchase verification. While legacy blue ticks have largely been removed at the time of writing, information on legacy verified status was still available from the Twitter API at the time of data collection, and these accounts mainly include public or institutional profiles with a large following.

\section{Results}
\subsection{Baseline Amplification Analysis}

Figure 2 illustrates the findings from the baseline amplification analysis, showing the mean percentage difference obtained from each fold of the bias corrected and accelerated bootstrapping (BCa) procedure \cite{diciccio}. Here, results reveal that on average, across 1000 stratified bootstrapping samples stratified by engagement level and followers count,  samples of low-credibility tweets generate more impressions than high-credibility samples across both datasets, with low-credibility tweets on COVID-19 receiving a baseline impressions amplification of +19.2\% (median +17.3\%) and low-credibility tweets on climate change generating +95.8\% impressions (median = +90.1\%). In absolute values, this amounts to a mean difference of +113.7 impressions (median = +111.4) for COVID-19 tweets, and +474.6 impressions (median = +447.2) for climate change tweets. This observation suggests that at an aggregate level across 1000 balanced samples of high-credibility and low-credibility tweets, the latter experiences heightened amplification within these two issue domains, with greater amplification in the context of climate change, indicating that, in aggregate, Twitter users were more likely to be exposed to low-credibility information by the platform’s recommender system. Results here also show that this behaviour is observed quite consistently, as 84.4\% of COVID-19 samples have a positive mean difference, and 97.9\% of climate change samples have a positive mean difference, suggesting that it is very rare that low-credibility samples will outperform high-credibility samples in impressions counts. 

\begin{figure*}[h!]
  \centering
  \includegraphics[width=0.7\textwidth]{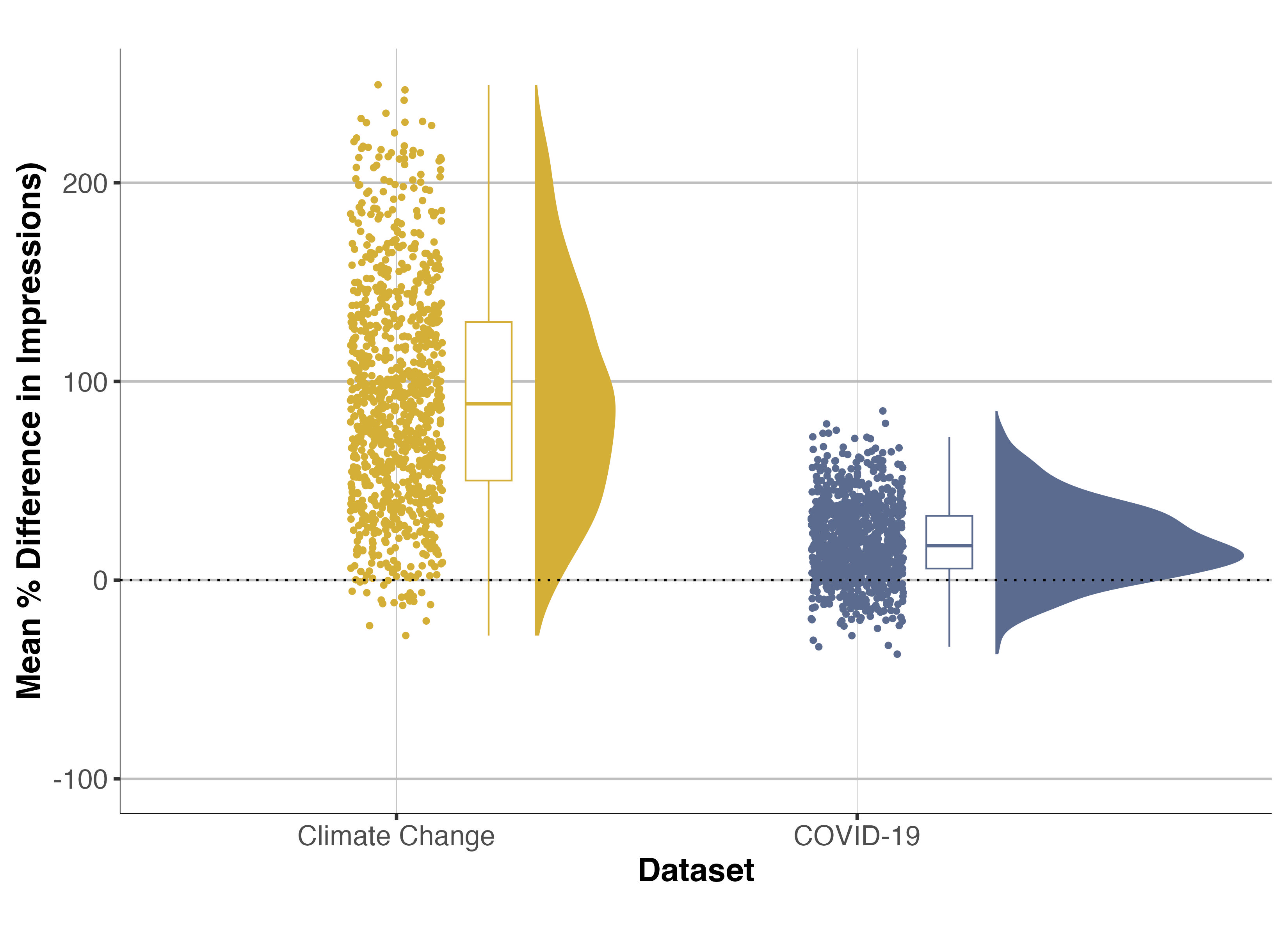}
  \caption{Raincloud plot illustrating the average percentage difference in impressions between high-credibility and low-credibility tweets, based on 1,000 resamples from each dataset under study.
}
  \label{fig:1}
\end{figure*}
         
However, when dealing with skewed distributions such as those of social media impressions, looking at the aggregate mean may not be sufficient to fully explain an amplification effect. Rather, we must also assess inter-stratum breakdowns of variabilities, which are shown in Figure 3, containing heatmaps of the mean differences in impressions across all 16 strata combinations as well as the size of each stratum. This step of the analysis delivers a more nuanced understanding of the results, showing that within bootstrapped samples, the observed difference in impressions is primarily generated by a difference in the highest-engagement and highest-followers strata (3,3). For COVID-19, this amounts to a mean difference of +3,148 impressions between low-credibility and high-credibility tweets, while for the climate change dataset, this amounts to +9,197. While the percentage of amplification within these strata appears extensive in absolute terms, it is more moderate in relative terms — showing +30.2\% amplification for the COVID-19 stratum 3.3 and +129\% for the climate change data within the same stratum. Qualitatively assessing tweets from these strata, they appear to largely be conspiratorial tweets from large right and far-right outlets. For example, the low-credibility tweet with the highest impressions in such strata cites an article from the Daily Mail, and states:  "A shadowy Army unit secretly spied on British citizens who criticised govt’s Covid lockdown policies… artificial intelligence deployed to 'scrape' social media for keywords".

This finding on the distribution of  intra-stratum variabilities is crucial for the interpretation of the bootstrapping results, as it shows that  while on aggregate users on Twitter were more likely to be exposed to low-credibility content, this effect is largely attributable to a difference in high-engagement, high-followers tweets, which are very impactful in terms of impressions generation, and are more likely receive amplified visibility when containing low-credibility content. This is consequential, as this minority of tweets are responsible for a large share of impressions generated on Twitter. This finding also aligns with the broader understanding of Twitter activity dynamics, where the majority of engagement and impressions are typically generated by a limited number of highly engaging posts, adding a further layer of understanding that this subgroup of posts is likely to achieve greater amplification when containing low-credibility content. 

\begin{figure*}[h!]
  \centering
  \includegraphics[width=0.7\textwidth]{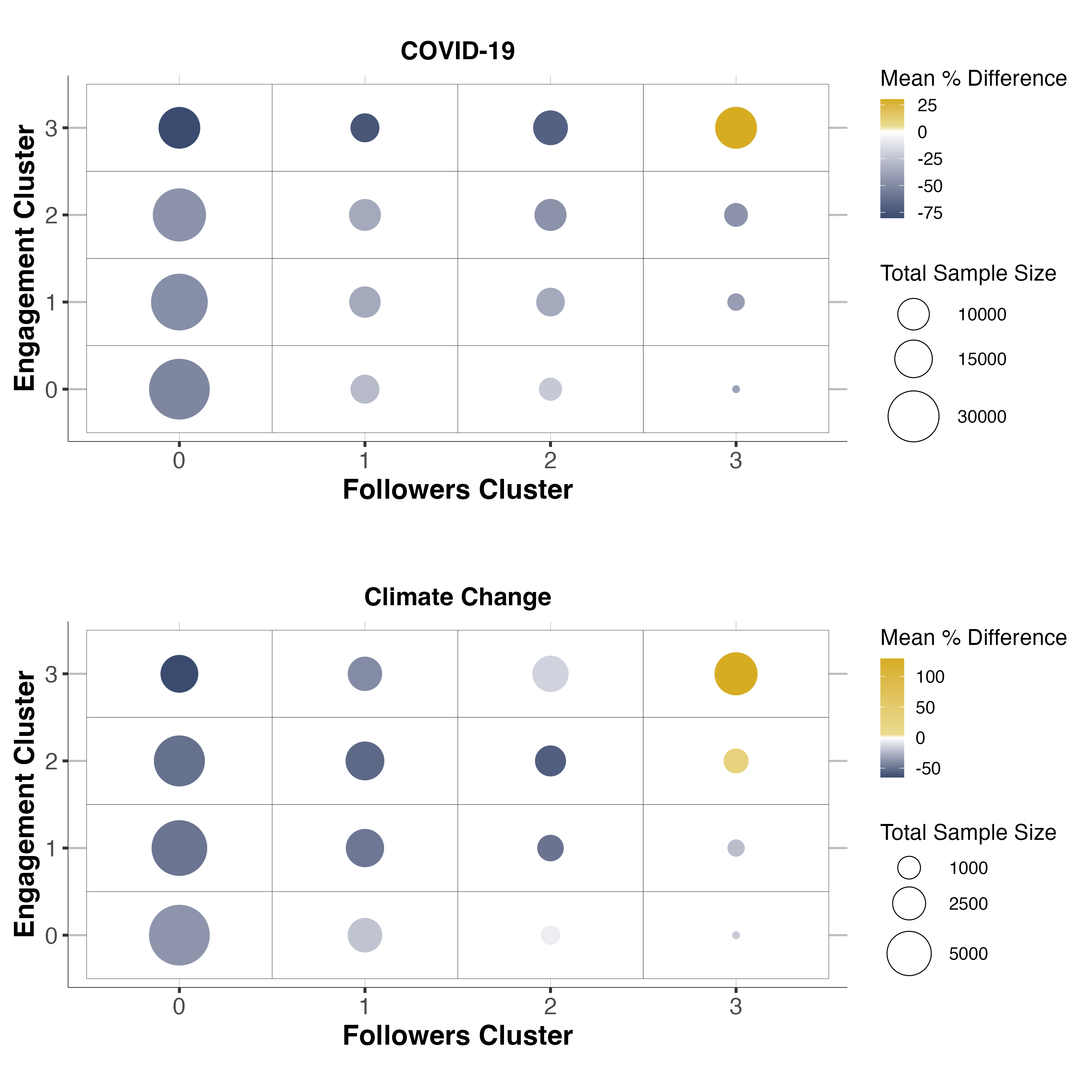}
  \caption{Heatmaps illustrating the percentage difference between low-credibility and high-credibility tweets in each of the 16 strata under analysis 
}
  \label{fig : 2}
\end{figure*}
         
\subsection{Analysis of Additional Stratifications}

Building on these observations, Figure 4 illustrates the impact of adding additional stratification variables - in this case toxicity, political bias and verified status -  to the amplification levels observed in the base model. Here, results showcase the average raw change of adding an additional stratification variable to the percentage amplification value observed in the baseline model. This step provides several insights into how additional stratification variables impact the amplification of low-credibility content. 

The first stratification variable added to the baseline model is a tweet’s toxicity profile, which provides insights into how the presence of inflammatory language influences the algorithmic amplification of low-credibility content. Examples of high-toxicity phrasing include strong insults, profanity, threats, and intentionally harmful or misleading rhetoric. These types of inflammatory expressions tend to appear in conspiracy theories and highly controversial claims, precisely the types of questionable information examined in this study. Here, results show no consistent relationship between amplification and toxicity for low and medium toxicity tweets across both the climate change and COVID-19 datasets. However, high-toxicity tweets exhibit a heightened algorithmic reach versus the baseline model for both datasets, at a value of  +9.9 and +15.2 respectively. This is important, as it suggests that tweets containing overtly negative, rude or disrespectful language see greater visibility on the platform when containing false or misleading domains. By showing a relationship between high-toxicity and algorithmic amplification across both datasets, these results support the understanding that content that is emotionally charged, especially when having a negative or controversial nature, may benefit from algorithmic amplification within engagement-based recommender systems. 

Furthermore, the addition of political bias as a stratification variable provides insights on the role of political partisanship in recommender-based amplification. Here, results show that for both COVID-19 and climate change datasets, tweets expressing right-leaning political bias see heightened amplification compared to the baseline model, and compared to left-leaning tweets. For COVID-19, right-leaning shows an increase of +6 from the baseline for low-credibility content versus high-credibility content, compared o a change of +0.03 for left-wing tweets. For climate change, right-leaning tweets exhibit an additional amplification of +24.3, compared to a value of +6 for left-leaning tweets. These results indicate that in the moment this analysis was carried out, tweets containing domains with a right-leaning political bias were more likely to be amplified in discussions on both COVID-19 and climate change, particularly if compared with tweets with a left-wing bias. This finding emerges as a crucial aspect of the current analysis, as it suggests that the Twitter recommender system, in its pursuit of maximising user engagement, may inadvertently expose users to content characterised by right-wing political biases. This observation gains particular significance in the context of ongoing debates surrounding filter bubbles, polarised information ecosystems, and the potential consequences of such dynamics on the public sphere. As the Twitter recommender system warrants increased visibility to content with pronounced right-wing political leanings, there is a clear risk that users may be driven towards more radical viewpoints - a trend that has been previously identified in Youtube’s recommender system \cite{haroon2022,ribeiro2020} - exacerbating existing social and political divisions while undermining the platform’s capacity to serve as a space for diverse and open discourse.

Lastly, to warrant more in-depth insights into user-level factors contributing to the performance of low-credibility tweets, the third additional layer of stratification is the verification status of a tweet’s author. Here, results clearly indicate that low-credibility tweets from users with a legacy verified status obtained evident amplification, with a change over the baseline model of +155\% for COVID-19 data, and + 138\% for climate change data. In contrast, tweets from unverified authors see minimal change compared to baseline amplification levels. This effect is very large, particularly if compared with the previous two stratifications, and indicates that low-credibility information spread by verified users is far more likely to be amplified on Twitter compared to low-credibility content shared by non-verified users, suggesting that the legacy checkmark, which acted as a credibility signal within the algorithm, may have been weaponised to amplify the reach of false or misleading content. This finding has important implications regarding the role of status cues and authority in algorithmic amplification, as it demonstrates that peripheral credibility signals like the verification status can override actual content quality in terms of driving engagement and reach. 

 \begin{figure*}[h!]
  \centering
  \includegraphics[width=0.7\textwidth]{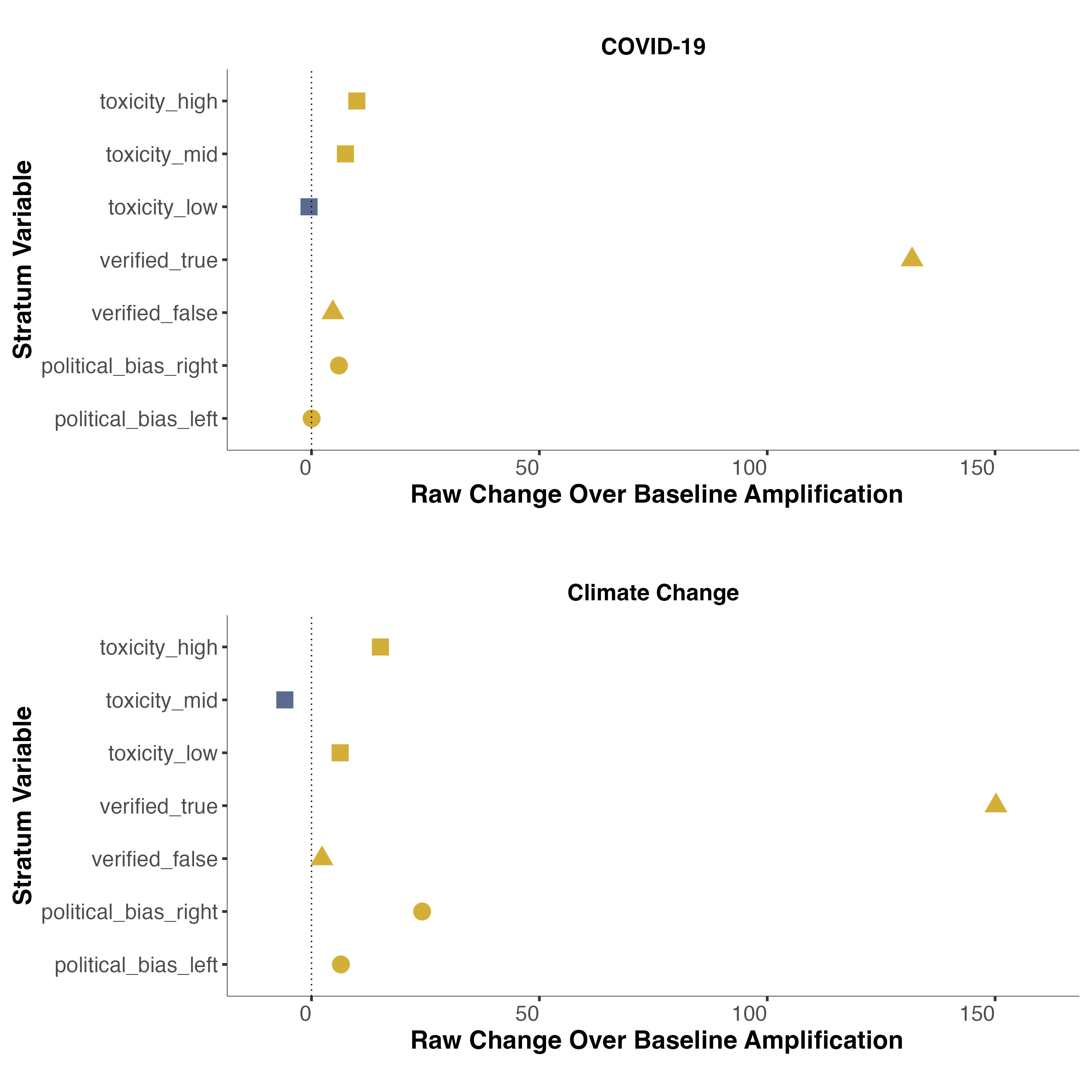}
  \caption{Raw impact of each value of additional stratifications on the baseline model of low-credibility amplification. Here, raw values mean that if in the base model a stratum had an amplification of +10\%, and the addition of a third stratification pushed this to +12\%, we would report this as a +2\% difference, not +20\%. In this sense, results here are a raw difference between percentages.}
  \label{fig: 3}
\end{figure*}
        
\section{Discussion and Conclusions}

The recent addition of impressions data on the Twitter API offers a unique opportunity to investigate the role of Twitter’s recommender system in promoting the circulation of disinformation and misinformation. While this opportunity has now been limited by restrictions to the Twitter API, this study presents an initial attempt to use an inferential approach based on impressions data to assess any differences between low-credibility and high-credibility tweets on Twitter. The main analysis of this work revealed that tweets containing URLs from low-credibility domains achieve higher visibility on the platform, with an average difference after bootstrapping of +19.2\% for COVID-19 data, and +95.8\% for climate change data. However, results also show that within bootstrapped samples, the majority of this effect comes from low-credibility tweets’ overperformance for high-engagement and high-followers users, which account for the majority of Twitter’s impressions. 

This work also set out to uncover notable features that may explain any impressions-based amplification, and found that toxicity scores obtained from Jigsaw’s Perspective API showed a clear pattern where high-toxicity tweets exhibited heightened amplification when containing low-credibility content, confirming the existing understanding that toxic content may be more easily amplified by engagement-based recommender systems. These results support findings from prior works, particularly those showing clear connections between negative emotional content and viral misinformation spread \cite{brady2017}. By confirming the role of engagement-based recommender systems in amplified toxic content, these findings highlight the need for alternative approaches to social media recommendation, such as bridging-based recommender systems, which focus on connecting users with diverse perspectives and high-quality information rather than maximising engagement alone\cite{ovadyathor}. While developing recommender systems that balance engagement, equality and diversity is indeed challenging, and further research is needed on this topic, this works lends support to the notion that alternatives to engagement-based recommendations are worth pursuing to improve online discourse and reduce the amplification of misinforming, toxic content.

Furthermore, one of the most notable results from the additional stratification analysis is the substantial amplification observed for politically biased low-credibility tweets, especially those with right-leaning partisanship. This finding highlights growing concerns about political polarisation and the rise of filter bubbles on social platforms. It suggests that in its pursuit of maximising engagement, the version of Twitter’s recommender system under analysis may have inadvertently contributed to echo chambers by disproportionately amplifying tweets that expressed partisan viewpoints, regardless of their truthfulness. This is particularly concerning given the vulnerability of high-profile socio-political discussions like climate change or COVID-19 to manipulation by misinformation campaigns. The fact that right-leaning tweets see substantially more amplification of false claims on climate change points to the ability of partisan disinformation to successfully exploit algorithmic loopholes on Twitter. This highlights concerns about social media deepening societal divides by siloing users into hyper-partisan circles, limiting exposure to alternate viewpoints. Lastly, results showed that low-credibility tweets from legacy verified accounts enjoy significant amplification on Twitter. While legacy verified accounts are only a minority of the platform’s population, it is important to note that Twitter recently changed its verification policy, and the blue tick can now be purchased as part of a monthly subscription plan. While data on paid-for verified users is not available via the Twitter API, this finding is concerning, as it suggests that it is possible that the blue tick, which warrants amplified visibility within the recommender system, may be weaponised by malicious actors to spread false and misleading information, particularly during highly emotionally loaded emergencies such as that of COVID-19. This point is crucial, and deserves further attention in future research.

Finally, this research offers initial empirical evidence on the promotion of low-credibility content on Twitter, revealing key factors that may contribute to a significant impressions overperformance of tweets containing false or misleading information. Given ongoing concerns about Twitter's impacts on information quality and integrity under its new management, this historical assessment of recommendation patterns can also offer an important benchmark for future analysis. Moreover, the study's methodological approach of leveraging impression counts to empirically analyse recommendation outcomes remains broadly relevant. Evaluating the visibility afforded to different types of content on social platforms is key to understanding how algorithmic curation shapes online information landscapes. By offering a framework to probe the role of recommender systems in propagating low-credibility information, this research retains value for inspiring further investigation - both of Twitter's evolving dynamics and other influential networks. Capturing past system behaviours can provide a foundation to monitor emerging risks and work towards safer, more accountable social information ecosystems.

Recommender systems are arguably the most prevalent application of AI and machine learning globally, affecting billions of internet users daily. However, despite their ubiquitous nature, there is an evident absence of regulation on their large-scale deployment. The results of this study highlight a need, as a minimum, for increased transparency in the field of social media recommender systems, as only through system-wide access it will be possible to produce clearer causal explanations for the effects observed in this work. However, as the recent release of parts of the Twitter algorithm eloquently shows, in principle transparency is not enough. Rather, it is crucial that transparency is accompanied the possibility of full replication of a recommender system,which would allow for the definition of protocols for comprehending, auditing, and stress-testing these systems to prevent the unintentional promotion of disinformation, to curb the perpetuation of harmful biases, and to safeguard the security of knowledge dissemination and acquisition processes.

Addressing this challenge is admittedly difficult, as recommender systems are proprietary, high-value assets at the heart of social media platforms. Nevertheless, it is vital to stress the significance of transparency and oversight for systems that mediate access to information and shape public discourse on a global scale. This study contributes to emphasising these imperatives, while also offering initial insights into how Twitter’s recommender system may have specifically amplified misleading and false information. Future research building on these findings should further investigate the dynamics of disinformation circulation on social media and explore potential interventions to curb the unintentional spread of misinformation by recommender systems. Ultimately, continued research in this domain is essential to ensure the responsible design and implementation of AI systems.

\printbibliography

@article{alatawi2021,
   author = {Alatawi, Faisal and Cheng, Lu and Tahir, Anique and Karami, Mansooreh and Jiang, Bohan and Black, Tyler and Liu, Huan},
   title = {A survey on echo chambers on social media: Description, detection and mitigation},
   journal = {arXiv preprint arXiv:2112.05084},
   year = {2021},
   type = {Journal Article}
}

@article{anandhan2018,
   author = {Anandhan, Anitha and Shuib, Liyana and Ismail, Maizatul Akmar and Mujtaba, Ghulam},
   title = {Social media recommender systems: review and open research issues},
   journal = {IEEE Access},
   volume = {6},
   pages = {15608-15628},
   ISSN = {2169-3536},
   year = {2018},
   type = {Journal Article}
}

@inproceedings{bhadani,
   author = {Bhadani, Saumya},
   title = {Biases in Recommendation System},
   booktitle = {Proceedings of the 15th ACM Conference on Recommender Systems},
   pages = {855-859},
   type = {Conference Proceedings}
}

@article{brady2017,
   author = {Brady, William J and Wills, Julian A and Jost, John T and Tucker, Joshua A and Van Bavel, Jay J},
   title = {Emotion shapes the diffusion of moralized content in social networks},
   journal = {Proceedings of the National Academy of Sciences},
   volume = {114},
   number = {28},
   pages = {7313-7318},
   ISSN = {0027-8424},
   year = {2017},
   type = {Journal Article}
}

@phdthesis{brennen2020,
   author = {Brennen, J Scott and Simon, Felix M and Howard, Philip N and Nielsen, Rasmus Kleis},
   title = {Types, sources, and claims of COVID-19 misinformation},
   university = {University of Oxford},
   year = {2020},
   type = {Thesis}
}

@article{cinelli2021,
   author = {Cinelli, Matteo and De Francisci Morales, Gianmarco and Galeazzi, Alessandro and Quattrociocchi, Walter and Starnini, Michele},
   title = {The echo chamber effect on social media},
   journal = {Proceedings of the National Academy of Sciences},
   volume = {118},
   number = {9},
   pages = {e2023301118},
   ISSN = {0027-8424},
   year = {2021},
   type = {Journal Article}
}

@article{falkenberg2022,
   author = {Falkenberg, Max and Galeazzi, Alessandro and Torricelli, Maddalena and Di Marco, Niccolò and Larosa, Francesca and Sas, Madalina and Mekacher, Amin and Pearce, Warren and Zollo, Fabiana and Quattrociocchi, Walter},
   title = {Growing polarization around climate change on social media},
   journal = {Nature Climate Change},
   pages = {1-8},
   ISSN = {1758-678X},
   year = {2022},
   type = {Journal Article}
}

@book{giansiracusa2021,
   author = {Giansiracusa, Noah},
   title = {How Algorithms Create and Prevent Fake News: Exploring the Impacts of Social Media, Deepfakes, GPT-3, and More},
   publisher = {Springer},
   ISBN = {1484271548},
   year = {2021},
   type = {Book}
}

@article{graham2020,
   author = {Graham, Timothy and Bruns, Axel and Zhu, Guangnan and Campbell, Rod},
   title = {Like a virus: The coordinated spread of coronavirus disinformation},
   year = {2020},
   type = {Journal Article}
}

@article{haroon2022,
   author = {Haroon, Muhammad and Chhabra, Anshuman and Liu, Xin and Mohapatra, Prasant and Shafiq, Zubair and Wojcieszak, Magdalena},
   title = {YouTube, The Great Radicalizer? Auditing and Mitigating Ideological Biases in YouTube Recommendations},
   journal = {arXiv preprint arXiv:2203.10666},
   year = {2022},
   type = {Journal Article}
}

@article{islam2019,
   author = {Islam, Rashidul and Keya, Kamrun Naher and Pan, Shimei and Foulds, James},
   title = {Mitigating demographic biases in social media-based recommender systems},
   journal = {KDD (Social Impact Track)},
   year = {2019},
   type = {Journal Article}
}

@article{kaiser2021,
   author = {Kaiser, Jonas and Rauchfleisch, Adrian and Córdova, Yasodara},
   title = {Comparative approaches to mis/disinformation| fighting zika with honey: An analysis of youtube’s video recommendations on brazilian youtube},
   journal = {International Journal of Communication},
   volume = {15},
   pages = {19},
   ISSN = {1932-8036},
   year = {2021},
   type = {Journal Article}
}

@article{lada2021,
   author = {Lada, Akos and Wang, Meihong and Yan, Tak},
   title = {How does news feed predict what you want to see},
   journal = {Personalized ranking with machine learning. Retrieved},
   volume = {28},
   pages = {2021},
   year = {2021},
   type = {Journal Article}
}

@article{leerssen2020,
   author = {Leerssen, Paddy},
   title = {The Soap Box as a Black Box: Regulating transparency in social media recommender systems},
   journal = {European Journal of Law and Technology},
   volume = {11},
   number = {2},
   year = {2020},
   type = {Journal Article}
}

@article{liu2022,
   author = {Liu, Zhuoran and Zou, Leqi and Zou, Xuan and Wang, Caihua and Zhang, Biao and Tang, Da and Zhu, Bolin and Zhu, Yijie and Wu, Peng and Wang, Ke},
   title = {Monolith: Real Time Recommendation System With Collisionless Embedding Table},
   journal = {arXiv preprint arXiv:2209.07663},
   year = {2022},
   type = {Journal Article}
}

@book{pariser2011,
   author = {Pariser, Eli},
   title = {The filter bubble: How the new personalized web is changing what we read and how we think},
   publisher = {Penguin},
   ISBN = {1101515120},
   year = {2011},
   type = {Book}
}

@article{pentina2014,
   author = {Pentina, Iryna and Tarafdar, Monideepa},
   title = {From “information” to “knowing”: Exploring the role of social media in contemporary news consumption},
   journal = {Computers in human behavior},
   volume = {35},
   pages = {211-223},
   ISSN = {0747-5632},
   year = {2014},
   type = {Journal Article}
}

@article{pierri2022,
   author = {Pierri, Francesco and DeVerna, Matthew R and Yang, Kai-Cheng and Axelrod, David and Bryden, John and Menczer, Filippo},
   title = {One year of COVID-19 vaccine misinformation on Twitter},
   journal = {arXiv preprint arXiv:2209.01675},
   year = {2022},
   type = {Journal Article}
}

@article{resnick2018,
   author = {Resnick, Paul and Ovadya, Aviv and Gilchrist, Garlin},
   title = {Iffy quotient: A platform health metric for misinformation},
   journal = {Cent Soc Media Responsib},
   volume = {17},
   pages = {1-20},
   year = {2018},
   type = {Journal Article}
}

@inproceedings{ribeiro2020,
   author = {Ribeiro, Manoel Horta and Ottoni, Raphael and West, Robert and Almeida, Virgílio AF and Meira Jr, Wagner},
   title = {Auditing radicalization pathways on YouTube},
   booktitle = {Proceedings of the 2020 conference on fairness, accountability, and transparency},
   pages = {131-141},
   year = {2020},
   type = {Conference Proceedings}
}

@article{santos2021,
   author = {Santos, Fernando P and Lelkes, Yphtach and Levin, Simon A},
   title = {Link recommendation algorithms and dynamics of polarization in online social networks},
   journal = {Proceedings of the National Academy of Sciences},
   volume = {118},
   number = {50},
   pages = {e2102141118},
   ISSN = {0027-8424},
   year = {2021},
   type = {Journal Article}
}

@misc{shearer2017, 
  title={News consumption across social media in 2017}, 
  author={Shearer, Elisa and Gottfried, Jeffrey}, 
  year={2017}, 
  howpublished={\url{https://www.pewresearch.org/journalism/2017/09/07/news-use-across-social-media-platforms-2017/}}, 
  note={Accessed: 20-Apr-2023}
}

@article{treen2020,
   author = {Treen, Kathie M d'I and Williams, Hywel TP and O'Neill, Saffron J},
   title = {Online misinformation about climate change},
   journal = {Wiley Interdisciplinary Reviews: Climate Change},
   volume = {11},
   number = {5},
   pages = {e665},
   ISSN = {1757-7780},
   year = {2020},
   type = {Journal Article}
}

@misc{walker2021, 
  title={News consumption across social media in 2021}, 
  author={Walker, Mason and Matsa, Katerina Eva}, 
  year={2021}, 
  howpublished={\url{https://www.pewresearch.org/journalism/2021/09/20/news-consumption-across-social-media-in-2021/}}, 
  note={Accessed: 20-Apr-2023}
}

@inproceedings{zhao,
   author = {Zhao, Zhe and Hong, Lichan and Wei, Li and Chen, Jilin and Nath, Aniruddh and Andrews, Shawn and Kumthekar, Aditee and Sathiamoorthy, Maheswaran and Yi, Xinyang and Chi, Ed},
   title = {Recommending what video to watch next: a multitask ranking system},
   booktitle = {Proceedings of the 13th ACM Conference on Recommender Systems},
   pages = {43-51},
   type = {Conference Proceedings}
}

@misc{thorburn_2022, 
  title={How platform recommenders work}, 
  author={Thorburn, Luke}, 
  year={2022}, 
  month={Nov},
  publisher={Understanding Recommenders},
  howpublished={\url{https://medium.com/understanding-recommenders/how-platform-recommenders-work-15e260d9a15a}}, 
  note={Accessed: 20-Apr-2023}
}

@misc{burns2023,
  title = {What TikTok’s Secret Heating Button Reveals About Virality Online},
  author = {Burns,Peter},
  howpublished = {\url{https://medium.com/feedium/algorithmic-heating-virality-is-a-choice-and-the-game-is-rigged-150307f1032a}},
  year = {2023},
}

@article{wang2021,
   author = {Wang, Zhiqiang and She, Qingyun and Zhang, Junlin},
   title = {MaskNet: Introducing feature-wise multiplication to CTR ranking models by instance-guided mask},
   journal = {arXiv preprint arXiv:2102.07619},
   year = {2021},
   type = {Journal Article}
}

@misc{twitterengineering,
  title = {Using Deep Learning at Scale in Twitter’s Timelines},
  author = {Twitter},
  howpublished = {\url{https://blog.twitter.com/engineering/en_us/topics/insights/2017/using-deep-learning-at-scale-in-twitters-timelines}},
  year = {2017},
}

@misc{twitteralgo,
  title = {The Twitter Algorithm: TweepCred},
  author = {Twitter},
  howpublished = {\url{https://github.com/twitter/the-algorithm/blob/main/src/scala/com/twitter/graph/batch/job/tweepcred/README}},
  year = {2023},
}

@article{Barrie,
   author = {Barrie, Christopher and Ho, Justin Chun-ting},
   title = {academictwitteR: an R package to access the Twitter Academic Research Product Track v2 API endpoint},
   journal = {Journal of Open Source Software},
   volume = {6},
   number = {62},
   pages = {3272},
   ISSN = {2475-9066},
   year = {2021},
   type = {Journal Article}
}

@inproceedings{Pfeffer,
   author = {Pfeffer, Juergen and Matter, Daniel and Jaidka, Kokil and Varol, Onur and Mashhadi, Afra and Lasser, Jana and Assenmacher, Dennis and Wu, Siqi and Yang, Diyi and Brantner, Cornelia},
   title = {Just another day on Twitter: a complete 24 hours of Twitter data},
   booktitle = {Proceedings of the International AAAI Conference on Web and Social Media},
   volume = {17},
   pages = {1073-1081},
   ISBN = {2334-0770},
   type = {Conference Proceedings}
}

@article{bovet2019,
   author = {Bovet, Alexandre and Makse, Hernán A},
   title = {Influence of fake news in Twitter during the 2016 US presidential election},
   journal = {Nature communications},
   volume = {10},
   number = {1},
   pages = {7},
   ISSN = {2041-1723},
   year = {2019},
   type = {Journal Article}
}

@article{kouzy2020,
   author = {Kouzy, Ramez and Abi Jaoude, Joseph and Kraitem, Afif and El Alam, Molly B and Karam, Basil and Adib, Elio and Zarka, Jabra and Traboulsi, Cindy and Akl, Elie W and Baddour, Khalil},
   title = {Coronavirus goes viral: quantifying the COVID-19 misinformation epidemic on Twitter},
   journal = {Cureus},
   volume = {12},
   number = {3},
   ISSN = {2168-8184},
   year = {2020},
   type = {Journal Article}
}

@article{lin2023,
   author = {Lin, Hause and Lasser, Jana and Lewandowsky, Stephan and Cole, Rocky and Gully, Andrew and Rand, David G and Pennycook, Gordon},
   title = {High level of correspondence across different news domain quality rating sets},
   journal = {PNAS Nexus},
   pages = {pgad286},
   ISSN = {2752-6542},
   year = {2023},
   type = {Journal Article}
}

@book{efrontibi94,
   author = {Efron, Bradley and Tibshirani, Robert J},
   title = {An introduction to the bootstrap},
   publisher = {CRC press},
   ISBN = {0412042312},
   year = {1994},
   type = {Book}
}

@article{jung2019,
   author = {Jung, Kwanghee and Lee, Jaehoon and Gupta, Vibhuti and Cho, Gyeongcheol},
   title = {Comparison of bootstrap confidence interval methods for GSCA using a Monte Carlo simulation},
   journal = {Frontiers in psychology},
   volume = {10},
   pages = {2215},
   ISSN = {1664-1078},
   year = {2019},
   type = {Journal Article}
}

@article{narayanan2023,
   author = {Narayanan, Arvind},
   title = {Understanding Social Media Recommendation Algorithms},
   year = {2023},
   type = {Journal Article}
}

@inbook{dougherty,
   author = {Dougherty, James and Kohavi, Ron and Sahami, Mehran},
   title = {Supervised and unsupervised discretization of continuous features},
   booktitle = {Machine learning proceedings 1995},
   publisher = {Elsevier},
   pages = {194-202},
   year = {1995},
   type = {Book Section}
}

@inproceedings{lees,
   author = {Lees, Alyssa and Tran, Vinh Q and Tay, Yi and Sorensen, Jeffrey and Gupta, Jai and Metzler, Donald and Vasserman, Lucy},
   title = {A new generation of perspective api: Efficient multilingual character-level transformers},
   booktitle = {Proceedings of the 28th ACM SIGKDD Conference on Knowledge Discovery and Data Mining},
   pages = {3197-3207},
   type = {Conference Proceedings}
}

@inproceedings{saveski,
   author = {Saveski, Martin and Roy, Brandon and Roy, Deb},
   title = {The structure of toxic conversations on Twitter},
   booktitle = {Proceedings of the Web Conference 2021},
   pages = {1086-1097},
   type = {Conference Proceedings}
}

@article{cuthbertson,
   author = {Cuthbertson, Lana and Kearney, Alex and Dawson, Riley and Zawaduk, Ashia and Cuthbertson, Eve and Gordon-Tighe, Ann and Mathewson, Kory W},
   title = {Women, politics and Twitter: Using machine learning to change the discourse},
   journal = {arXiv preprint arXiv:1911.11025},
   year = {2019},
   type = {Journal Article}
}

@article{petter,
   author = {Törnberg, Petter},
   title = {Chatgpt-4 outperforms experts and crowd workers in annotating political twitter messages with zero-shot learning},
   journal = {arXiv preprint arXiv:2304.06588},
   year = {2023},
   type = {Journal Article}
}

@article{gilardi2023,
   author = {Gilardi, Fabrizio and Alizadeh, Meysam and Kubli, Maël},
   title = {Chatgpt outperforms crowd-workers for text-annotation tasks},
   journal = {arXiv preprint arXiv:2303.15056},
   year = {2023},
   type = {Journal Article}
}

@article{diciccio,
   author = {Diciccio, Thomas J and Romano, Joseph P},
   title = {A review of bootstrap confidence intervals},
   journal = {Journal of the Royal Statistical Society Series B: Statistical Methodology},
   volume = {50},
   number = {3},
   pages = {338-354},
   ISSN = {1369-7412},
   year = {1988},
   type = {Journal Article}
}

@misc{thorburnetal2023,
   author = {Thorburn, Luke and  Stray, Jonathan and Bengani, Priyanjana},
   title = {Making Amplification Measurable},
   url = {https://medium.com/understanding-recommenders/making-amplification-measurable-2be548e5986c},
   year = {2023},
   type = {Online Multimedia}
}

@article{ovadyathor,
   author = {Ovadya, Aviv and Thorburn, Luke},
   title = {Bridging Systems: Open Problems for Countering Destructive Divisiveness across Ranking, Recommenders, and Governance},
   journal = {arXiv preprint arXiv:2301.09976},
   year = {2023},
   type = {Journal Article}
}

@misc{newsguard,
   author = {Newsguard},
   title = {Score and Rating Levels},
   url = {https://www.newsguardtech.com/ratings/rating-process-criteria/},
   year = {2023},
   type = {Web Page}
}
\end{document}